\documentclass[draft,11pt]{article}
\usepackage[dvips]{epsfig}

 \advance \textwidth by -2in
 \advance \textheight by -3in
\def\##1{\underline #1}
\def\=#1{\underline{\underline #1}}

\def\epso{\epsilon_0}
\def\muo{\mu_0}
\def\ko{k_0}

\def\etao{\eta_0}
\def\.{\mbox{ \tiny{$^\bullet$} }}

\def\epsa{\epsilon_a}
\def\epsar{\epsilon'_a}
\def\epsai{\epsilon''_a}

\def\mua{\mu_a}
\def\muar{\mu'_a}
\def\muai{\mu''_a}

\def\epsb{\epsilon_b}
\def\mub{\mu_b}

\def\ux{\#{u}_x}
\def\uy{\#{u}_y}
\def\uz{\#{u}_z}

\def\les{\left[}
\def\ris{\right]}
\def\lec{\left\{}
\def\ric{\right\}}

\def\r#1{(\ref{#1})}

\begin{document}

\noindent {\Large {\bf On Planewave Remittances and Goos--H\"anchen Shifts
of Planar Slabs with Negative Real Permittivity and Permeability}}
\vskip 0.2cm

\noindent  {\bf AKHLESH LAKHTAKIA\/}
\vskip 0.2cm
\noindent {CATMAS~---~Computational \& Theoretical Materials Sciences Group\\
Department of Engineering Science \& Mechanics\\
Pennsylvania State University, University Park, PA 16802--6812, USA}

\bigskip

\noindent  {\bf Abstract:} {\em Two results
about the planewave response of an isotropic,
dielectric--magnetic,
homogeneous, planar slab are deduced when the real parts of both the
permittivity and the permeability are altered from positive to negative. First,
 the reflection and the transmission coefficients suffer phase reversals
without change in their respective magnitudes. Second, the Goos--H\"anchen
shifts experienced by beams on total reflection reverse their  directions.
}

\bigskip


\bigskip

\section{Introduction}
The recent publication of an experimental realization of certain
artificial materials~---~that are effectively  isotropic, homogeneous,
and possess negative real permittivity and permeability in some
frequency range (Shelby, Smith, \& Schultz, 2001)~---~has created
a big splash in the electromagnetics research community. Although the
first samples of these materials are quite primitive (Lakhtakia, 2001),
there is enough experimental as well as theoretical evidence (Dewar, 2001;
McCall, Lakhtakia, \& Weiglhofer, 2001)
along with the promise of applications (Pendry, 2001), that concerted
theoretical research in advance of its experimental counterpart is
justified.

This communication focuses on two aspects of the planewave
responses of planar slabs whose permittivity and permeability
have negative real parts at the frequency of interest. The first aspect comprises the magnitudes and phases
of the reflection and the transmission coefficients; and
the second encompasses the longitudinal shifts experienced by beams on
total reflection. An $\exp(-i\omega t)$ time--dependence is
implicit, with $\omega$ as the angular frequency.

\section{Theoretical Preliminaries}
Suppose the region $0 \leq z \leq L$ is occupied by an
isotropic, homogeneous, dielectric--magnetic medium
with relative permittivity $\epsa=\epsar+i\epsai$ and
relative permeability $\mua=\muar+i\muai$. The medium
being passive, the principle of energy conservation requires that
$\epsai \geq 0$ and $\muai\geq 0$. The half--spaces
$z < 0$ and $z>L$ are vacuous. A plane wave is incident
on the face $z=0$ of the slab; accordingly, a plane wave
is reflected into the half--space $z<0$ and another plane
wave is transmitted into the half--space $z>L$.

\subsection{s--polarization}
Let the incident plane wave be $s$--polarized. Then, the
electric field phasor in the two half--spaces can be
expressed as follows:
\begin{equation}
\label{eq1}
{\bf E}(z) =
\uy \exp(i\kappa x)\,\lec
\begin{array}{ll}
 \exp(i\ko\cos\theta_{inc}z) + r_s\,\exp(-i\ko\cos\theta_{inc}z) 
\,, & \;\; z\leq 0\\
t_s \exp\les i\ko\cos\theta_{inc}(z-L)\ris \,, & \;\; z\geq L
\end{array}
\right.\,.
\end{equation}
Here and hereafter, $\ko=\omega\sqrt{\epso\muo}$ is the wavenumber in vacuum;
$\epso$ and $\muo$ are the permittivity and the
permeability of vacuum, respectively; $\etao=\sqrt{\muo/\epso}$
is the intrinsic impedance
of vacuum; the
quantity
\begin{equation}
\kappa =\ko\,\sin\theta_{inc}\,
\end{equation}
involves $\theta_{inc}$ as the angle of incidence measured from the $+z$ axis;
$\ux$, $\uy$ and $\uz$ are cartesian unit vectors;
while $r_s$ and $t_s$ are the reflection and the transmission
coefficients, respectively.

Following standard practice, and after defining the 2$\times$2 matrixes
\begin{equation}
\label{s-defmp}
[P_s] =\les\begin{array}{cc}
0 & -\omega\muo\mua \\-\omega\epso\epsa +\frac{\kappa^2}{\omega\muo\mua} & 0
\end{array}\ris\,,\quad
[M_s] = \exp\lec i [P_s]\,L\ric\,,
\end{equation}
we obtain the following matrix equation for the boundary
value problem described in the previous two paragraphs:
\begin{equation}
\label{s-sys}
t_s\,
\les\begin{array}{c}
 1\\ -\,\etao^{-1}\, \cos\theta_{inc} \end{array}\ris
=[M_s]\,
\les\begin{array}{c}
r_s +1 \\ (r_s-1)\,\etao^{-1}\, \cos\theta_{inc}\end{array}\ris\,.
\end{equation}
This equation's solution yields $r_s$ and $t_s$, from which
the remittances $R_s=\vert r_s\vert^2$ and $T_s=\vert t_s\vert^2$ 
can be calculated.

\subsection{p--polarization}
If the incident plane wave is $p$--polarized, the magnetic
field phasor outside the slab $0\leq z\leq L$ can be written
as
\begin{equation}
{\bf H}(z) =
-\,\uy\,\etao^{-1}\,\exp(i\kappa x)\,\lec
\begin{array}{ll}
 \exp(i\ko\cos\theta_{inc}z) + r_p\,\exp(-i\ko\cos\theta_{inc}z) 
\,, & \;\; z\leq 0\\
t_p \exp\les i\ko\cos\theta_{inc}(z-L)\ris \,, & \;\; z\geq L
\end{array}\right.
\,.
\end{equation}
The solution of the boundary value problem involves the calculation
of the coefficients $r_p$ and $t_p$ from the matrix
equation
\begin{equation}
\label{p-sys}
-\,t_p\,
\les\begin{array}{c}
 \cos\theta_{inc}\\ \etao^{-1}\end{array}\ris
=[M_p]\,
\les\begin{array}{c}
(r_p -1)\,\cos\theta_{inc}\\ -\,(r_p+1)\,\etao^{-1}\end{array}\ris\,,
\end{equation}
where
\begin{equation}
\label{p-defmp}
[M_p]= \exp\lec i [P_p]\,L\ric\,,\quad
[P_p]= \les\begin{array}{cc}
0 &\omega\muo\mua-\frac{\kappa^2}{\omega\epso\epsa}\\
\omega\epso\epsa & 0\end{array}\ris\,.
\end{equation}
The remittances $R_p=\vert r_p\vert^2$ and $T_p=\vert t_p\vert^2$
can be obtained thereafter.

\section{Reflection and Transmission  Analysis}
Both matrix equations \r{s-sys} and \r{p-sys}
can be stated together compactly as
\begin{equation}
\label{sys}
t\, [f] + r\, [M]\,[g] = [M]\,[h]\,, \quad [M] = \exp\lec i[P]\,L\ric\,,
\end{equation}
where $[f]$, $[g]$ and $[h]$ are 2$\times$1
matrixes.

Now, $\epsar > 0$ and $\muar>0$ if the region $0
\leq z\leq L$ is filled with a conventional medium.
Suppose the transformation 
\begin{equation}
\label{tfm1}
\lec\epsar\,\rightarrow\,-\,\epsar\,,\,\muar\,\rightarrow\,-\,\muar\ric\,
\end{equation}
occurs; i.e., the conventional medium is replaced by a counterpart
the real parts of whose permittivity and permeability are exactly
negative of those of the conventional medium, but both $\epsai$
and $\muai$ remain unchanged.

\subsection{Magnitude Invariance and Phase Reversal}
According to \r{s-defmp} and
\r{p-defmp}, the transformation
\begin{equation}
\label{tfm2}
\lec [P]\,\rightarrow\,-\,[P]^\ast\ric\,
\end{equation}
must follow, with the asterisk denoting the complex conjugate. As
shown in the Appendix, we get
\begin{equation}
\label{tfm3}
\lec [M]\,\rightarrow\,\,[M]^\ast\ric\,
\end{equation}
in consequence. The matrixes $[f]$, $[g]$ and $[h]$ in
\r{sys} being real--valued,  \r{tfm3} in turn mandates
the transformation
\begin{equation}
\label{tfm4}
\lec r\,\rightarrow\,r^\ast,\, t\,\rightarrow\,t^\ast\ric \,
\end{equation}
for the reflection and the transmission coefficients,
and
\begin{equation}
\label{tfm5}
\lec R\,\rightarrow\,R\, , T\,\rightarrow\,T\ric \,
\end{equation}
for the remittances.

The symmetry inherent in the
relationship \r{tfm1}$\Rightarrow$\r{tfm4} is of
significance for experiments on planar, isotropic dielectric--magnetic
slabs.
When the signs of the real parts of both the
permittivity and the permeability of a medium are altered,
the planewave reflection and transmission coefficients remain
unchanged in magnitude but suffer a phase reversal.
The alteration cannot be detected if only the planewave remittances
are measured.

\subsection{Goos--H\"anchen Shifts}
A major consequence of the relationship
\r{tfm1}$\Rightarrow$\r{tfm4} can be appreciated when the incident
plane wave is replaced by a beam. By definition,
a plane wave has infinite transverse extent; but a beam
is of finite transverse extent.  Newton
had conjectured that if a beam were to impinge on a
planar interface with an electromagnetically rarer medium and total reflection were
to occur, the reflected beam would emerge with a longitudinal displacement
(along the $+x$ axis). Such
displacements are often called the Goos--H\"anchen shifts after the two
scientists who reported the first measurements in 1947
(Goos \& H\"anchen, 1947). 
Subsequently, these shifts have been estimated as well as measured for planar
interfaces between many different pairs
of materials (Lakhtakia, 2002; Lotsch, 1970a,b, 1971a,b),
 and
several applications have emerged
as well (Jackson, 1999; Chauvat et al., 2000; de Fornel, 2000).

Analytical treatment of the Goos--H\"anchen shifts {\em essentially}
requires the solution of the same problem as dealt with
in Section 2 (Haibel, Nimtz,  \&   Stahlhofen, 2001). 
The vacuous half--spaces are, however,
filled with a medium possessing $\epsb$ and $\mub$
as its relative permittivity and relative permeability,
both positive and  real--valued. Equations \r{eq1}--\r{p-defmp}
still apply, but with (i) $\ko$ replaced by $\ko\sqrt{\epsb\mub}$,
and (ii) $\etao$ by $\etao\sqrt{\mub/\epsb}$. The relationship
\r{tfm1}$\Rightarrow$\r{tfm4} still holds.

\bigskip
\begin{center}
--------------------------------------------------------------------------------------------------------
\bigskip
\begin{figure}[!ht]
\centerline{ \psfull
\epsfig{file=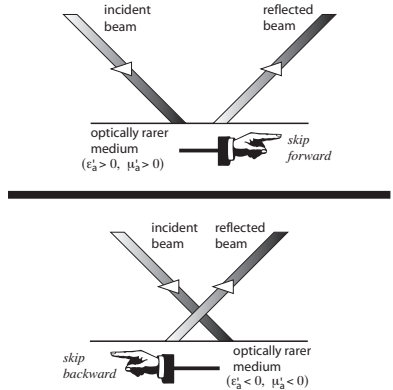,width=2in}} 

\caption{Schematics for Goos--H\"anchen shifts.}
\end{figure}
\bigskip
--------------------------------------------------------------------------------------------------------\end{center}
\bigskip

With the assumption of a Gaussian profile for the incident beam,
and after denoting the reflection coefficient $r = \vert r\vert\, \exp(i\varphi)$,
the longitudinal shift is estimated
as
\begin{equation}
\label{def-d}
d=-\,\frac{\partial\varphi}{\partial\kappa}\,.
\end{equation}
This quantity is meaningful, in the present
context, only when $\theta_{inc}$ exceeds
the critical angle; i.e., when total
reflection occurs. A depth of penetration equal to $(d/2)\cot\theta_{inc}$
is also sometimes calculated.

Two Goos--H\"anchen shifts are possible in the present scenario:
$d_s$ and $d_p$ for $s$-- and $p$--polarized beams,
respectively. Their
characteristics are well--known (Jackson, 1999; de Fornel, 2000)
and do not bear repetition here. It suffices to note
here that 
\begin{equation}
\label{GH}
\lec\epsar\,\rightarrow\,-\,\epsar\,,\,\muar\,\rightarrow\,-\,\muar\ric\,\Rightarrow
\lec d_s\,\rightarrow\,-\, d_s,\,d_p \,\rightarrow\,-\, d_p\ric\,,
\end{equation}
this relationship being an immediate consequence 
of
the relationship
\r{tfm1}$\Rightarrow$\r{tfm4} and the definition \r{def-d}.

The result \r{GH} can be understood 
with the help of Figure 1 as follows: When total reflection
occurs and the electromagnetically rarer medium is of the conventional
type (i.e., $\epsar>0$ and $\muar>0$), the reflected beam appears
as if the incident beam has skipped forward. On the contrary,
according to \r{GH}, the reflected beam must appear to skip
backward, if the electromagnetically rarer medium is of the kind
reported by Shelby, Smith, \& Schultz (2001) (i.e., 
$\epsar<0$ and $\muar<0$). No violation of causality
would occur in the latter instance, because some time
must elapse during the traversal of the distance 
that is apparently skipped. This result is also
consistent with the propagation velocity being oppositely
directed to the energy velocity when $\epsar<0$ and $\muar<0$
(McCall, Lakhtakia, \& Weiglhofer, 2001).

\section{Conclusions}

Examining the alteration in the planewave response of an isotropic,
dielectric--magnetic,
homogeneous, planar slab when the real parts of both the
permittivity and the permeability are changed from positive to negative,
we arrived at the following deductions:
\begin{itemize}
\item[(i)]
 The planewave reflection and transmission coefficients suffer phase reversals.
\item[(ii)] 
The planewave remittances remain unaffected.
\item[(iii)] When conditions for total reflection prevail, the longitudinal
shift experienced by a beams reverses its direction.
\end{itemize}
\bigskip

\section*{Appendix}
Let a matrix $[P]$ of size $m\times m$
be diagonalizable. Then it can be decomposed
as 
\begin{equation}
[P] = [U]\,[G]\,[U]^{-1}\,,
\end{equation}
where the matrix $[G]={\rm diag}(g_1,\,g_2,\cdots \,g_m)$ contains the 
consecutive eigenvalues
of $[P]$, while the successive columns of $[U]$ are the corresponding
eigenvectors of $[P]$. It follows that (Hochstadt, 1975)
\begin{equation}
\exp\lec i[P]z\ric = [U]\,\exp\lec i[G]z\ric\,[U]^{-1}\,.
\end{equation}

Let the matrix $[Q]=-[P]^\ast$ so that
\begin{equation}
[Q] = [U]^\ast\,\Big(-[G]^\ast\Big)\,\Big([U]^{-1}\Big)^\ast\,.
\end{equation}
Consequently,
\begin{eqnarray}
\nonumber
\exp\lec i[Q]z\ric &=& [U]^\ast\,\exp\Big(-i[G]^\ast z\Big)\,\Big([U]^{-1}\Big)^\ast
\\
\nonumber
&=& \Big([U]\,\exp\lec i[G]z\ric\,[U]^{-1}\Big)^\ast
\\
&=& \Big( \exp\lec i[P]z\ric\Big)^\ast\,.
\end{eqnarray}

\bigskip\bigskip\bigskip

\newpage

Chauvat, D., O. Emile, F. Bretenaker, \& A. Le Floch. 2000.
Direct measurement of the Wigner delay associated
with the Goos--H\"anchen effect.
{\em Phys. Rev. Lett.\/}  84:71--74.

de Fornel, F. 2000.
{\em Evanescent waves~---~From Newtonian optics to
atomic optics.\/} 
Berlin: Springer; pp. 12--18.

Dewar, G.A. 2001.
Candidates for $\mu < 0$, $\epsilon < 0$ nanostructures.
{\em Int. J. Mod. Phys. B\/} 15:3258--3265.

Goos, F. \& H. H\"anchen. 1947.
Ein neuer und fundamentaler Versuch zur Total\-reflexion.
{\em Ann. Phys. Lpz.\/} 1:333--346.

Haibel, A., G. Nimtz, \& A.A. Stahlhofen. 2001.
Frustrated total reflection: the double--prism revisited.
{\em Phys. Rev. E\/} 63:047601.

Hochstadt, H. 1975
{\em Differential equations~---~A modern approach.\/}
New York, NY: Dover Press; p. 57.

Jackson, J.D. 1999.
{\em Classical electrodynamics, 3rd ed.\/}
New York: Wiley; pp. 306--309.

Lakhtakia, A. 2001.
An electromagnetic trinity from ``negative permittivity"
and ``negative permeability".
{\em Int. J. Infrared Millim. Waves\/} 22:1731--1734.

Lakhtakia, A. 2002.
Truncation of angular spread of Bragg zones by total reflection,
and Goos--H\"anchen shifts
exhibited by chiral sculptured thin films.
{\em Arch. Elektr. \"Uber.\/} 56:000--000 (accepted
for publication).

Lotsch, H.K.V. 1970a.
Beam displacement at total reflection: The Goos--H\"anchen
effect, I.
{\em Optik\/} 32: 116--137.

Lotsch, H.K.V. 1970b.
Beam displacement at total reflection: The Goos--H\"anchen
effect, II.
{\em Optik\/} 32: 189--204.

Lotsch, H.K.V. 1971a.
Beam displacement at total reflection: The Goos--H\"anchen
effect, III.
{\em Optik\/} 32: 299--319.

Lotsch, H.K.V. 1971b.
Beam displacement at total reflection: The Goos--H\"anchen
effect, IV.
{\em Optik\/} 32: 553--569.

McCall, M.W., A. Lakhtakia, \& W.S. Weiglhofer. 2001.
The negative index of refraction demystified. Report No. 2001/30.
Department of Mathematics, University of Glasgow, UK.

Pendry, J. 2001.
Electromagnetic materials enter the negative age.
{\em Physics World\/} 14(9):47--51.

Shelby, R.A., D.R. Smith, \& S. Schultz. 2001.
Experimental verification of a negative index of refraction.
{\em Science\/} 292:77--79.

\end{document}